\pdfoutput=1
\documentclass{article}

\usepackage[final]{nips_2016}

\usepackage[utf8]{inputenc} % allow utf-8 input
\usepackage[T1]{fontenc}    % use 8-bit T1 fonts
\usepackage{hyperref}       % hyperlinks
\usepackage{url}            % simple URL typesetting
\usepackage{booktabs}       % professional-quality tables
\usepackage{amsfonts}       % blackboard math symbols
\usepackage{nicefrac}       % compact symbols for 1/2, etc.
\usepackage{microtype}      % microtypography
\usepackage[]{graphicx}
\usepackage{grffile}
\usepackage{amsmath}
\usepackage{verbatim}
\usepackage{subcaption}

\usepackage[font={small}]{caption}
\title{Identification of Cancer Patient Subgroups \\ via Smoothed Shortest Path Graph Kernel}

\author{
  Ali Burak \"Unal \\
  Department of Computer Science\\
  Bilkent University\\
  Ankara, Turkey \\
  \texttt{burak.unal@bilkent.edu.tr} \\
  \And
  \"Oznur Ta\c{s}tan \\
  Department of Computer Science\\
  Bilkent University\\
  Ankara, Turkey \\
  \texttt{oznur.tastan@cs.bilkent.edu.tr} \\
}

\begin{document}

\maketitle
\vspace{-1cm}
\begin{abstract}

Characterizing patient somatic mutations through next-generation sequencing technologies opens up possibilities for refining cancer subtypes. However, catalogues of mutations reveal that only a small fraction of the genes are altered frequently in patients. On the other hand different genomic alterations may perturb the same pathways. We propose a novel clustering procedure that quantifies the similarities of patients from their mutational profile on pathways via a novel graph kernel. We represent each KEGG pathway as an undirected graph. For each patient the vertex labels are assigned based on her altered genes. Smoothed shortest path graph kernel (smSPK) evaluates each pair of patients by comparing their vertex labeled pathway graphs. Our clustering procedure involves two steps: the smSPK kernel matrix derived for each pathway are input to kernel k-means algorithm and each pathway is evaluated individually. In the next step, only those pathways that are successful are combined in to a single kernel input to kernel k-means to stratify patients. Evaluating the procedure on simulated data showed that smSPK clusters patients up to 88\% accuracy.  Finally to identify ovarian cancer patient subgroups, we apply our methodology to the cancer genome atlas ovarian data that involves 481 patients.  The identified subgroups are evaluated through survival analysis. Grouping patients into four clusters results with patients groups that are significantly different in their survival times ($p$-value $\le 0.005$).

\end{abstract}
\vspace{-0.5cm}

\section{Introduction}
\vspace{-0.2cm}
%There is a pressing need for diagnostic, and therapeutic solutions for cancer. Potential solutions are hindered by difficulties in the effective classification of disease subtypes. 
Cancer is heterogeneous at the molecular level; seemingly similar tumors that are classified into the same histopathological cancer type may bear very different genomic alterations and follow very distinct clinical trajectories.  Recent decrease in cost and increase in resolution of sequencing technologies facilitated the accurate mapping of somatic mutations in large cohorts of cancer patients. This catalogue of genomic alterations opens up new opportunities for defining subtypes of many cancer types.

A major challenge in grouping patients based on their mutations is that there are very few genes that are altered frequently whereas there are many infrequently mutated genes  \cite{creixell2015pathway}. On the other hand, although different genes are affected in each patient, they may affect the same cellular mechanism. Therefore, instead of considering individual genes, mapping mutations on pathways and analyzing mutated pathways have been proposed as a more effective strategy for interpreting genetic alterations \cite{creixell2015pathway}. Pathways are diagrams that summarize interactions of well-studied cellular processes. These interactions represent regulation and signaling events, and biochemical reactions. In this work we present a novel approach where we assess patient similarities based on their mutational landscape on pathways through a novel graph kernel and use this kernel to cluster patients.

Graph kernels have been proposed in literature for measuring the similarity of pairs of graphs and are shown to be successful in comparing structured objects \cite{shervashidze2011weisfeiler, hido2009linear, borgwardt2005shortest}. The suggested graph kernels; however, are designed for comparing graphs with different topological structure and focus on finding similar subgraphs. In our problem, though, for each pathway we derive a single undirected graph, where the vertices represent genes and the edges represents interactions between genes. For each patient the graph vertices are labelled based on the set of mutations the patient harbors (Figure\ref{fig:examplepatients}). Thus, the graphs that we want to compare are topologically identical but the vertex label distributions are different. To this end, we propose a smoothed shortest path graph kernel (smSPK). smSPK compares two graphs based on their vertex label distribution taking into account the underlying graph topology.

Our clustering procedure involves comparing patients based on their mutational profiles on KEGG pathways using smSPK.  The stratification of patients in this way provides incorporating prior biological knowledge across different pathways and capture patients similarities that are arising due to dysregulation of similar process in the pathways. We first evaluated smSPK on synthethic data and this results confirm that smSPK can asses graphs similarities accurately. Next, we applied this procedure to find subgroup of ovarian cancers. Ovarian cancer is a major cause of cancer-related mortality in women and the standard treatment is aggressive surgery followed by chemotherapy \cite{jemal2010cancer}. It is one of the cancer types where identifying subgroups of patients is critically important \cite{cancer2011integrated}. Our results found clusters of patients. We used survival analysis to evaluate the clustering results. Below we first describe the data curation and processing steps, next the smoothed shortest path kernel, the clustering procedure. Finally we present our results.

\vspace{-0.3cm}
\section{Methods}
\vspace{-0.4cm}
\subsection{Data Curation and Processing }
\vspace{-0.2cm}
\noindent {\bf Cancer patient data: } Ovarian serous cystadenocarcinoma somatic mutation dataset is downloaded from Broad Institute GDAC Firehose \cite{mut}. These data originates from the Cancer Genome Atlas (TCGA) project and additionally includes the predicted functional impact of mutations \cite{reva2011predicting}. The functional impact is presented in four categories: neutral, low, medium, high. We disregard the mutations in the neutral category. The survival times and the survival status of the patients are directly retrieved from the TCGA data portal. The dataset includes 462 patients with mutational and clinical data available. These patients were mutated in 6851 different genes.

\vspace{-0.1cm}
\noindent{\bf Pathway data and preprocessing:} We download pathways from KEGG database \cite{kanehisa2015kegg} and parse them using KEGGParser \cite{arakelyan2013keggparser}. We process each pathway such that only genes are kept and nodes that represent entity types such as ortholog, map and others and their relations are discarded. We treat compounds differently. KEGG compounds are collections of molecules that are relevant to biological pathways such as lipids and sugars. Often times if they interact with a gene product, there are edges in and out of the compound. Thus, when removing {\it compound} type entries, we insert new edges between genes to which the compound is connected so that information flow is preserved. There are 264 pathways in total. Each pathway is converted to an undirected graph.

\vspace{-0.4cm}

\begin{figure}[ht]
\centering
\includegraphics[width=0.9\linewidth]{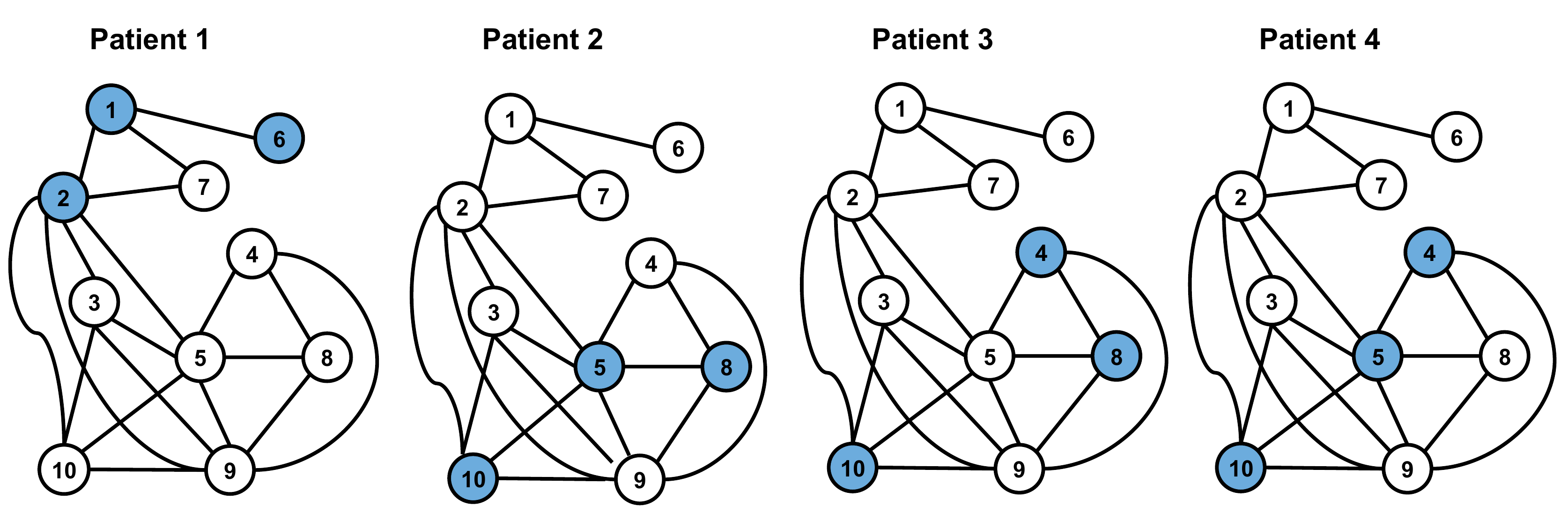}
\caption[Example undirected label graphs of patients on a single pathway]{ Mutational profiles of patients shown on an example undirected graph derived from the same pathway. Blue nodes indicate mutated genes and white nodes indicate unaltered genes. }
\label{fig:examplepatients}
\end{figure}

\vspace{-0.4cm}
\subsection{Smoothed Shortest Path Graph Kernel}
\vspace{-0.2cm}
For each pathway $i$ and patient $j$ we define an undirected vertex labeled graph $G_i^{(j)} = ( V_j,E, \ell )$. $V = \{v_1, v_2, \ldots, v_n\}$ is the ordered set of $n$ genes in the pathway and $E \subset V \times V$ is a set of undirected edges between genes. The label set $\ell = \{l_1, l_2, \ldots, l_n\}$ is in the same order of $V$ and represents the corresponding vertex's label. $l$ is assigned based on patient's mutational profile; if the corresponding gene is mutated in patient $i$, label $1$ is assigned and $0$ otherwise. Thus, graphs defined on the same pathway have the same topology but different label distribution. The adjacency matrix for the graph is $n \times n$ matrix ${\bf A}$ with ${\bf A}_{ij} = 1$ if there is an edge between $v_i$ and $v_j$, and 0 otherwise.

We would like to devise a kernel function that will compare pairs of graphs with the same topology but different label distribution. The graph kernel function should reflect the similarities in the label distribution of patients using the graph as the context. For instance, despite the set of nonidentical altered genes, if two patients have alterations in genes nearby close proximity, the should influence their similarities. We would also want the central nodes to have more influence compared to genes that are in the periphery of the pathway. For example, consider the four patients displayed in Figure\ref{fig:examplepatients}. Patient 1 is different from the rest of the patients because she harbors mutations on a different part of the pathway. The other three patients are similar to each other, as they have identical or close by altered genes. Among patients 2, 3 and 4, patient $2$ and patient $4$ bear the largest similarity as they are both mutated in a central gene (vertex 5) as opposed to a gene in the periphery of the pathway. 

Our suggested kernel function compares patient mutations along all the shortest paths of the undirected graph. In order the kernel function to consider changes in the neighbors, we first smooth the mutations along each shortest path. A shortest path for all patients is smoothed as follows \cite{hofree2013network}:

\vspace{-0.2cm}
\begin{equation}
{\bf S}_{t+1}= \alpha {\bf S}_{t} {\bf A} + \left(1 - \alpha \right) {\bf S}_{0}
\end{equation}

where $S_{0}$ is a patient-by-gene matrix which represents mutational states of the genes are on the shortest path and determined by $\ell$. If there is a mutation in that patent, the entry is 1 and if there is no mutation the entry of the matrix is 0. $S_{t}$ is the same matrix at time $t$. {\bf A} is the degree normalized adjacency matrix. $\alpha \in [0,1]$ is the parameter that defines the degree of smoothing. We iterate over propogation until convergence.

Let ${\bf s}_p^{(i)}$ be the vector that represents the smoothed mutational profile of patient $i$ on shortest path $p$. Let $N$ be the number of shortest path on a graph $g$. smoothed shortest path kernel for patients $i$ and $j$ for graph $g$ is defined as follows:

\vspace{-0.2cm}
\begin{equation}
\begin{aligned}
K_g (i,j)= \sum_{p=1}^N {\bf s}_p^{(i)} . ({\bf s}_p^{(j)})^T 
\end{aligned}
\end{equation}
The above function is a valid function, as the dot product is linear kernel and kernel property is preserved under summation.  When we apply this function on the example patients shown in   Figure\ref{fig:examplepatients},  patient1 is not similiar to the rest of the patients. Of the other three patients, highest similarity was among Patient 2 and 4 as we would like to achive.  Applying this function on all patient pairs, we obtain  ${\bf K}$ a patient-by-patient kernel matrix that reflects similarities of the patients on a single pathway. We compute these kernel matrices for each of the KEGG matrix and perform clustering as described in the next section. 

\vspace{-0.2cm}
\subsection{Clustering with Pathway Kernels}
\vspace{-0.2cm}
In the first step, we first compute kernel matrices on each KEGG pathway. Then we perform clustering of patients with each of these pathways. For a set $k$ value, we cluster the patients with kernel k-means with 100 random restarts \cite{shawe2004kernel}. In the next step, survival distributions of resulting patient clusters are compared using Kaplan-Meier survival curves and log-rank test. We disregard the pathways that fail to generate clusters that are not different in  terms of their survival distributions  based on the $p$-values of log-rank test. The remaining pathways' kernel matrices  are summed with uniform weight and are normalized.  To get the final clustering result, we apply kernel k-means on the combined kernel. The clustering result, then, is re-evaluated by survival analysis. We perform clustering by varying $k= 2,3,4,5$, the threshold $p$-value for filtering pathways and the smoothing parameter $\alpha$. 

%The outline of our approach is shown Figure \ref{fig:outline}.
%
%\begin{figure} [h]
%\centering
%\includegraphics[width=0.8\linewidth]{../figures/outline.pdf}
%\caption[a]{The overview of our approach}
%\label{fig:outline}
%\end{figure}

\vspace{-0.2cm}
\subsection{Synthetic Data Experiments}
\vspace{-0.2cm}
In order to assess smSPK, we generate synthetic patient data on a single pathway and perform kernel k-means clustering. We pick one pathway from KEGG with 45 genes and assume that there are 3 different subgroups with 200 patients in each. We further assume that for each cluster a different shortest path within the pathway is the driver and is mutated more often in comparison to other genes. We pick a shortest path for each group randomly such that its length is between the length of the longest shortest path and the half of it. Then for each subgroup we generate patient mutation data by randomly mutating genes on the shortest pathway with probability $p_{in}$. Additionally, for each patient we mutate the rest of the genes in the pathway with probability $p_{out}$. Thus in each cluster of patients the driver genes are concentrated on the same pathway but not on identical genes. Having clustered the patients, we calculate the accuracy as the percent of patients with correct cluster assignment. We repeat this experiment by varying $p_{in}$, $p_{out}$ and smoothing parameter $\alpha$. Each parameter configuration is repeated 100 times and average accuracy is calculated over these 100 simulations.

\vspace{-0.5cm}
\section{Results} 
\vspace{-0.3cm}
\begin{figure}
\centering
\includegraphics[width=0.5\linewidth]{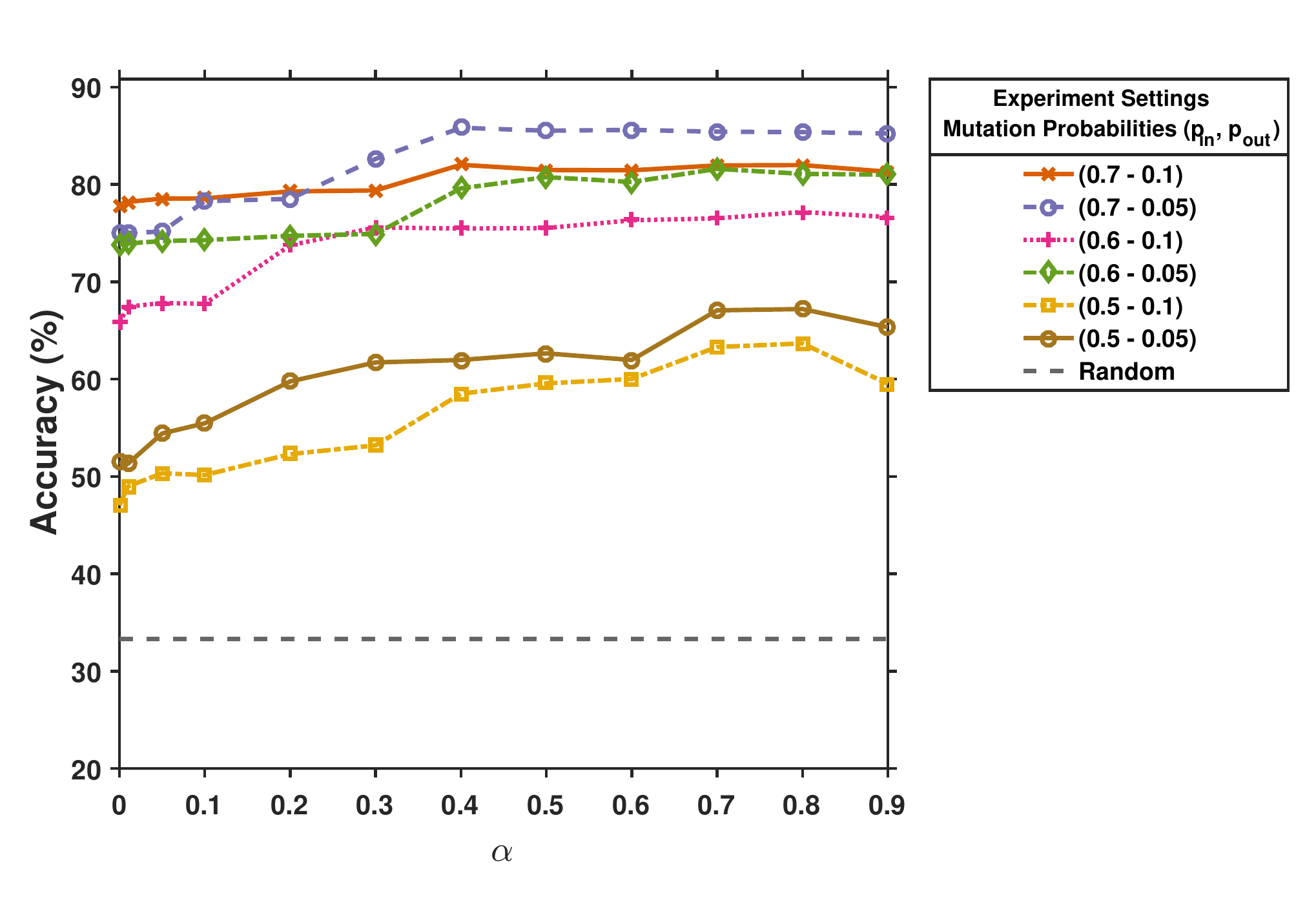}
\vspace{-0.2cm}
\caption[Results of synthetic data experiments]{Accuracy of clustering on synthetic mutation data on a 45 gene pathway. Three clusters with 200 patients are generated, where each cluster have a different path as a driver. For a patient, the genes on the cluster's driver path is mutated with probability $p_{in}$ and other genes in the pathway are mutated with probability $p_{out}$. The figure displays the average accuracy of cluster assignments over 100 repeated simulations for each parameter configuration. The x-axis varies the smoothing parameter $\alpha$. The dashed gray line is accuracy that will be achieved if the patients were randomly assigned to the three clusters.}
\label{fig:simulated}
\end{figure}

\begin{figure*}[t!]
\vspace{-0.7cm}
$\begin{array}{ccc}
\includegraphics[width=0.32\textwidth]{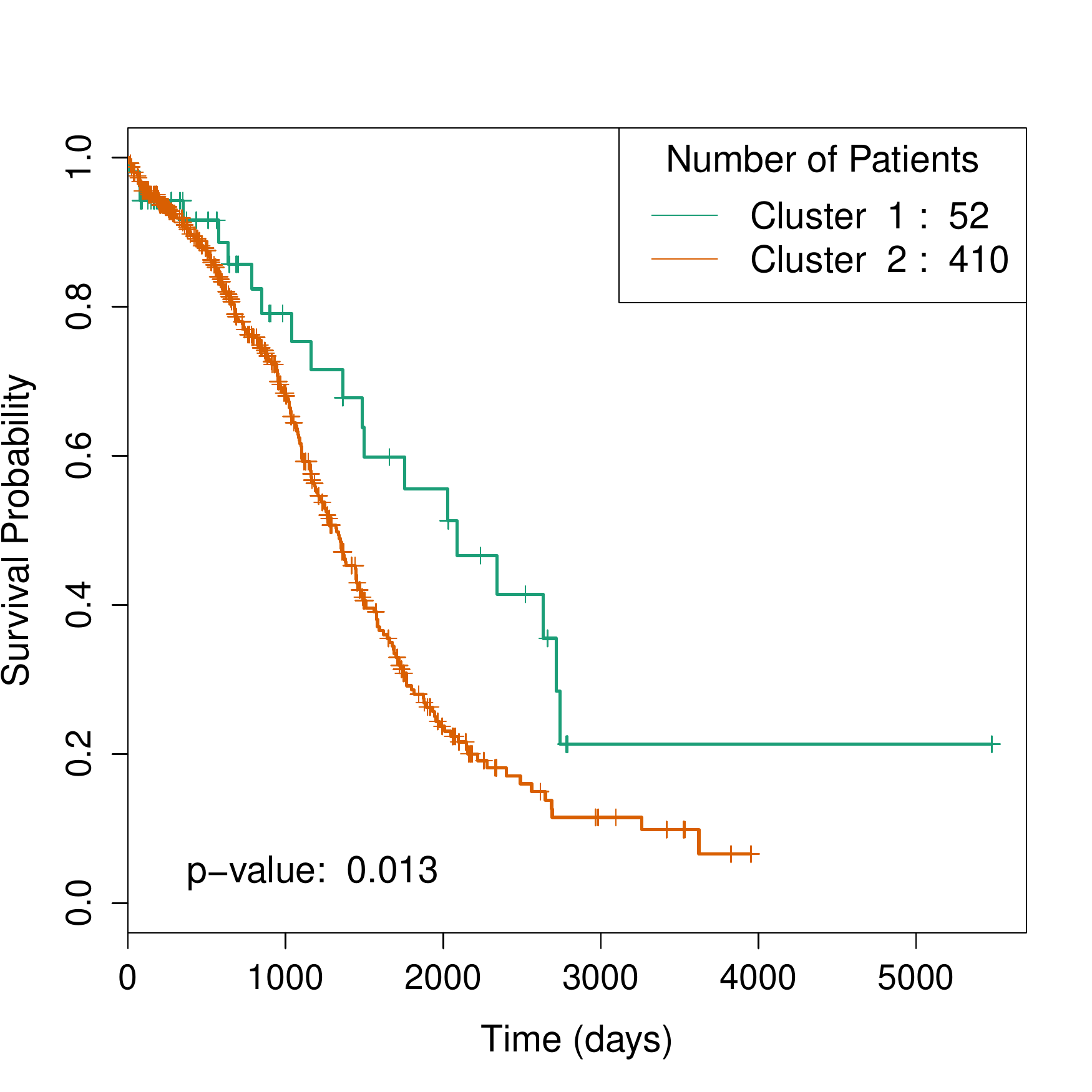} &
\includegraphics[width=0.32\textwidth]{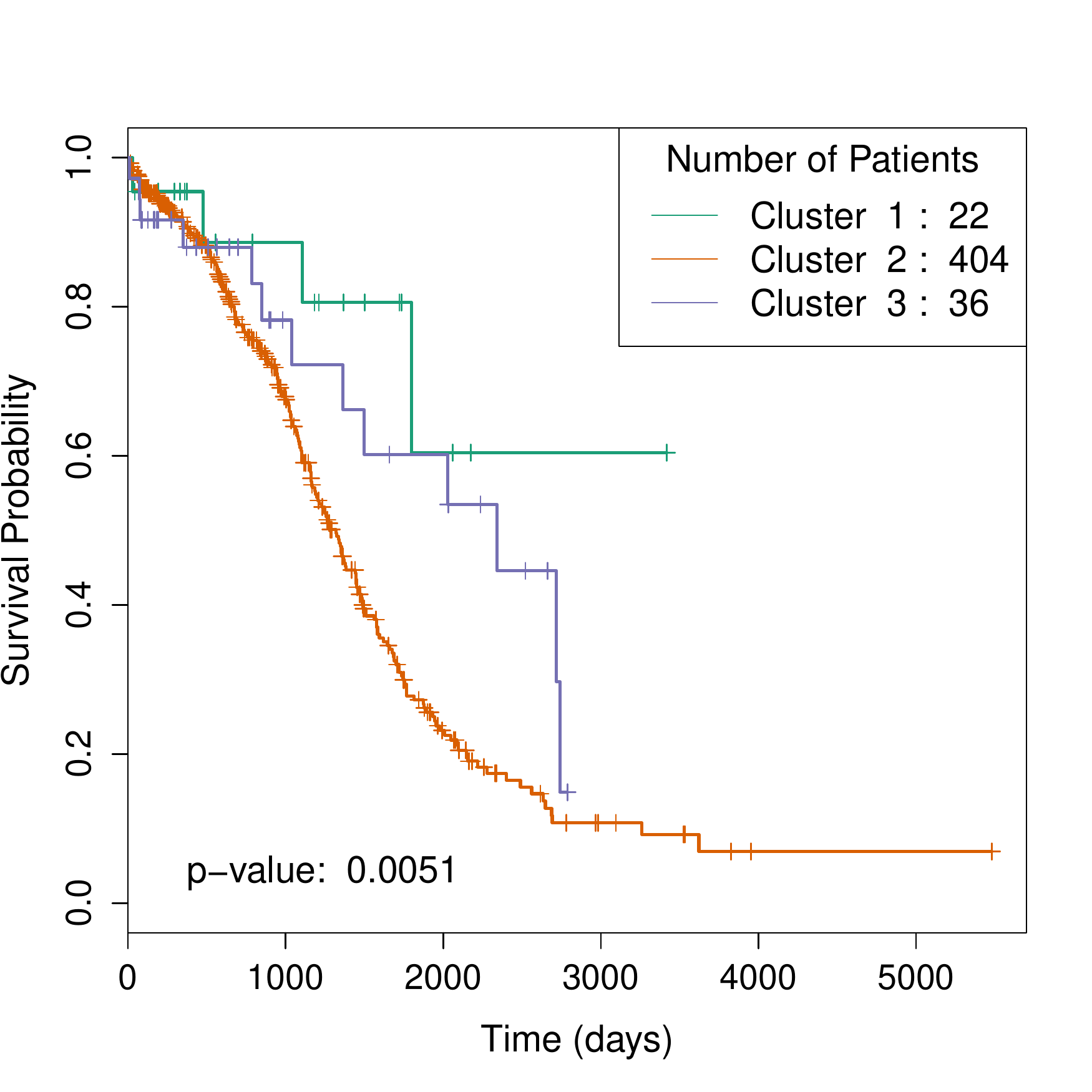} &
\includegraphics[width=0.32\textwidth]{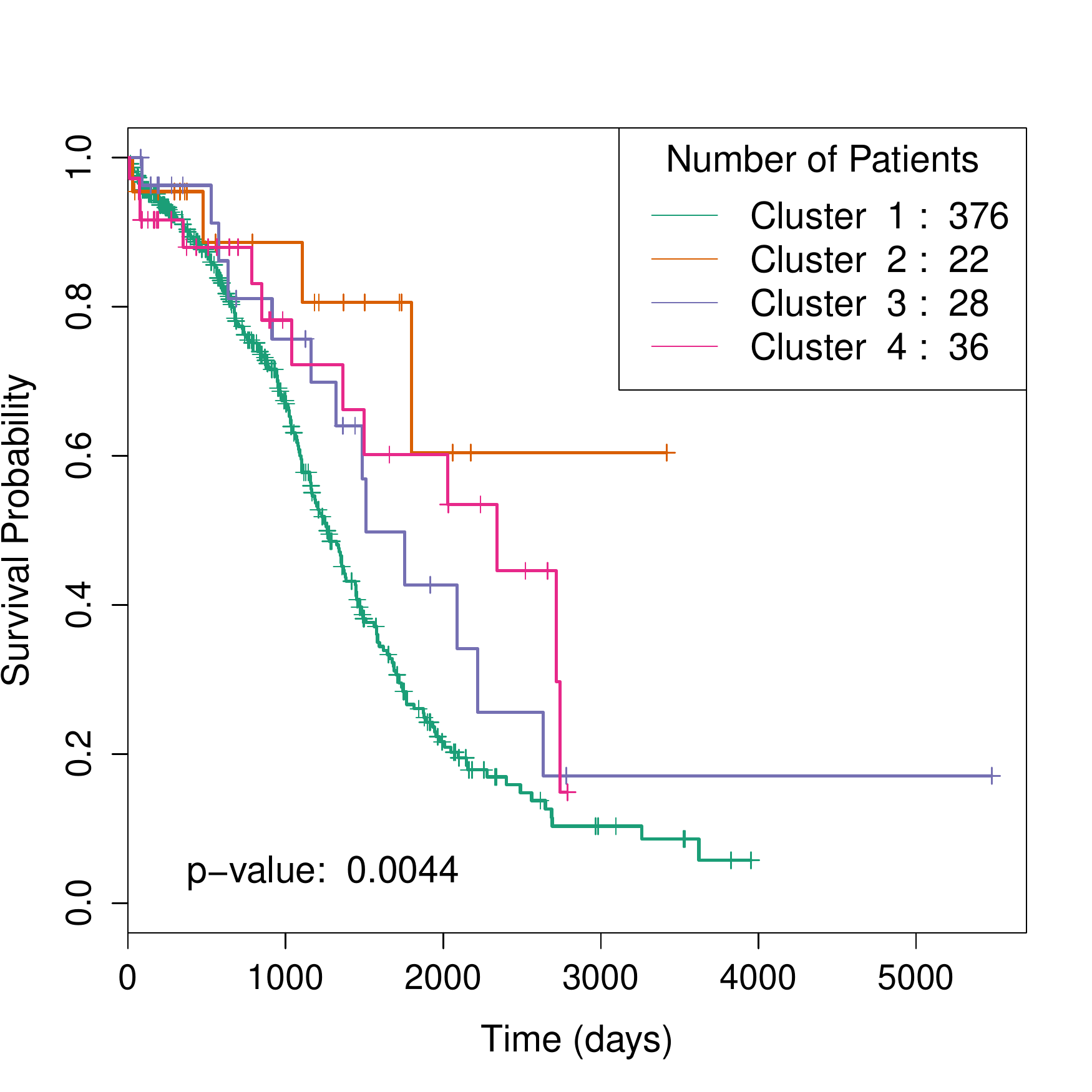}
\end{array}$
\vspace{-0.5cm}
\caption[Ovarian]{Kaplan-Meir survival plots of ovarian patient subgroups for different number of clusters (2, 3 and 4). The kernel matrices are calculated with smoothed shortest path kernel. The p-value is the result of log-rank test. }
\label{fig:ovarian}
\end{figure*}

Figure~\ref{fig:simulated} summarizes results  on synthetic data. We evaluated the clusters based on correct cluster assignment of patients. We experimented with different parameter configurations. The synthetic data experiments show that if the signal of subgroup is high enough, the stratification of patients can be achieved with very high accuracy up to 88 \%. They also reveal that if the mutations rate in non-driver genes is low, the accuracy is higher. $\alpha=0$  corresponds to case the path is not smoothed, as shown in graph smoothing on shortest paths improves the accuracies. Almost for all cases, spreading the effect of mutations through other genes in shortest path leads to more accurate separation of patients. Increasing $\alpha$ to some extent improves the results. 

We next applied our methodology to stratify ovarian cancer patients. We experimented with different parameter settings; kernel k-means $k$ value, threshold parameter for filtering each pathway and smoothing parameter $\alpha$ are varied. In each configuration, the clustering qualities are evaluated based on their success in finding subgroup of patients that differ in their survival distributions. Figure~\ref{fig:ovarian} displays the best results for $k=2, 3$ and $4$. As can be seen in all three results, cluster $1$ shows very distinct survival. Patients in this cluster have higher survival rates. Clustering patients into 3 or 4 clusters also reveal distinct groups. Evaluating based on $p$-value $k=3$ reveals the best results. We also evaluated clusters using kernelized silhouette width (data not shown) and observe the clusters are well seperated.  As a future work, we will be analyzing the driver pathways of each group as identifying subgroups of patients with similar genomic alterations can reveal unique molecular characteristics of these patient groups and opens up possibilities for targeted therapeutic regimens. 

Integrating other genomic and  transcriptomic data improves the understanding of a given cancer type. Our framework is very flexible in that we can easily incorporate changes in mRNA levels of genes or abundance of proteins measured.  As a second future work, the driver pathways that lead to clustering of ovarian patients and their relation to drivers reported in the literature will be explored.

%\begin{figure}[h]
%\begin{subfigure}{.5\textwidth}
%  \centering
%  \includegraphics[width=0.5\linewidth]{figure_4a.pdf
%  \caption{KM plot of experiment with $k = 2$}
%  \label{fig:sfig1}
%\end{subfigure}%
%\begin{subfigure}{.5\textwidth}
%  \centering
%  \includegraphics[width=0.9\linewidth]{figure_4a.pdf}
%  \caption{KM plot of experiment with $k = 3$}
%  \label{fig:sfig2}
%\end{subfigure}
%\begin{subfigure}{.5\textwidth}
%  \centering
%  \includegraphics[width=0.5\linewidth]{ovarian_sm_spk_alpha_0.8_threshold_0.01_k_4.pdf}
%  \caption{KM plot of experiment with $k = 4$}
%  \label{fig:sfig2}
%\end{subfigure}
%\caption{The resulting Kaplan-Meir survival plots of patients in different clusters. experiment in which we select pathways having lower p-value than $0.1$ and $0.01$ for consensus clustering respectively. In this experiment, we select $\alpha$ as $0.8$}
%\label{fovarian}
%\end{figure}

% **********************************************************
% **********************************************************
% **********************************************************

\bibliographystyle{unsrt}
\bibliography{mybib}{}

\end{document}